\documentclass[aps,prd,unsortedaddress,superscriptaddress,showpacs,nofootinbib,twocolumn,10pt]{revtex4-2} %notitlepage,
\usepackage{natbib}
\usepackage{booktabs}
\usepackage[colorlinks=true, linkcolor=blue, citecolor=blue, urlcolor=blue]{hyperref}
\usepackage[utf8]{inputenc}
\usepackage{multirow}
\usepackage{graphicx}
\usepackage{relsize}
\usepackage{slashed}
\usepackage[normalem]{ulem}
\usepackage{rotating}
\usepackage{bigstrut}
\usepackage{makecell}
\usepackage[dvipsnames]{xcolor}
\usepackage{amsmath}
\usepackage{amssymb,bm}
\usepackage{amsthm}
\usepackage{mathrsfs}
\usepackage{array}
\usepackage[all]{xy}
\usepackage{enumerate}
\usepackage{enumitem}
\usepackage{slashed}
\usepackage{float}
\usepackage{mathtools}
\usepackage{soul}
\usepackage{orcidlink}
\allowdisplaybreaks[4]

\newcommand{\ket}[1]{| #1 \rangle}

\definecolor{poscolor} {RGB} {252,188,190} 
\definecolor{negcolor} {RGB} {168,168,234} 
\usepackage{colortbl}

\usepackage{relsize}
\def\babar{\mbox{\slshape B\kern-0.1em{\smaller A}\kern-0.1em
    B\kern-0.1em{\smaller A\kern-0.2em R}}}

\usetikzlibrary{decorations.markings}
\usetikzlibrary{snakes}
\tikzset{
 photon/.style={decorate, decoration={snake}, draw=black},
    electron/.style={draw=black, postaction={decorate},
        decoration={markings,mark=at position .55 with {\arrow[draw=black]{>}}}},
    gluon/.style={decorate, draw=magenta,
        decoration={coil,amplitude=3pt, segment length=4pt}},
    scalar/.style={dashed,line width=.6pt, postaction={decorate}}
}

\newcommand{\itp}{\affiliation{Institute of Theoretical Physics, Chinese Academy of Sciences, Beijing 100190, China}}
\newcommand{\ucas}{\affiliation{School of Physical Sciences, 
University of Chinese Academy of Sciences, Beijing 100049, 
China}}

\newcommand{\peng}{\affiliation{Peng Huanwu Collaborative Center for Research and Education, International Institute for Interdisciplinary and Frontiers, Beihang University, Beijing 100191, China}}

\newcommand{\bonn}{\affiliation{Helmholtz Institut f\"{u}r Strahlen- und Kernphysik and Cluster of Excellence  ``Color meets Flavor'', Universit\"{a}t Bonn, D-53115 Bonn, Germany}}
\newcommand{\bethe}{\affiliation{Bethe Center for Theoretical Physics, Universit\"{a}t Bonn, D-53115 Bonn, Germany}}

\newcommand{\scnt}{\affiliation{Southern Center for Nuclear-Science Theory (SCNT), Institute of Modern Physics,\\ 
Chinese Academy of Sciences, Huizhou 516000, China}}

\newcommand{\fzj}{\affiliation{Institute for Advanced Simulation (IAS-4) and Cluster of Excellence ``Color meets Flavor'', Forschungszentrum J\"ulich, D-52425 J\"ulich, Germany}}

\begin{document}

\title{Precise determination of the properties of $X(3872)$ and of its isovector partner $W_{c1}$}
\author{Teng Ji\orcidlink{0000-0003-0366-1042}}
\bonn\bethe

\author{Xiang-Kun Dong\orcidlink{0000-0001-6392-7143}}
 \email[Corresponding author: ]{xiangkun@hiskp.uni-bonn.de}
\bonn\bethe

\author{Feng-Kun~Guo\orcidlink{0000-0002-2919-2064}}
\email[Corresponding author: ]{fkguo@itp.ac.cn}
\itp\ucas\scnt

\author{Christoph Hanhart\orcidlink{0000-0002-3509-2473}}\fzj

\author{Ulf-G. Mei{\ss}ner\orcidlink{0000-0003-1254-442X}}\bonn\bethe\fzj\peng

\begin{abstract}  
We perform a simultaneous fit to BESIII data on $e^+e^-\to \gamma (D^0{\bar{D}^{0}}\pi^0/J/\psi\pi^+\pi^-)$ and LHCb data on $B^+\to K^+(J/\psi\pi^+\pi^-)$ to precisely determine the properties of the $X(3872)$, with full consideration of three-body effects from $D^*\to D\pi$ decay, respecting both analyticity and unitarity. The $X(3872)$ is determined to be a {quasi-}bound state with a significance of $2.7\,\sigma$, representing the most precise determination to date. Its pole is located at    {$\left(-160^{+57}_{-74}-125^{+23}_{-38}\,i\right) \rm keV$}, relative to the nominal $D^0\bar{D}^{*0}$ threshold. 
Moreover, we confirm the presence of an isovector partner state, $W_{c1}$. It is found as a virtual state at  {$\left(3.1\pm0.7+ 1.3^{+1.9}_{-0.6}\,i\right)\ \rm MeV$} relative to the $D^+ D^{*-}$ threshold on an unphysical Riemann sheet, strongly supporting a molecular nature of both $X(3872)$ and $W_{c1}$. 
As a highly nontrivial prediction we show that the $W_{c1}$ leads to nontrivial lineshapes {around 3.88~GeV} in $B^0\to K^0X(3872)\to K^0 D^0\bar D^0\pi^0$ and $K^0J/\psi\pi^+\pi^-$---thus the scheme presented here can be tested further by improved measurements.

\bigskip
\noindent {\bf Key words}: Exotic hadrons, Hadronic molecules, Coupled-channel dynamics, Chiral effective field theory, $X(3872)$
\end{abstract}

\maketitle

\section{Introduction}

Since over 20 years the study of exotic hadronic states is one of the central themes in hadron physics. These states, which lie beyond the conventional quark model for quark-antiquark mesons and three-quark baryons, offer a unique opportunity to understand the inner workings of quantum chromodynamics (QCD), since
different multiquark configurations mean different realizations of confinement: in molecular states, confinement happens in the smallest possible subsystems, typically conventional hadrons, which are then bound together via a residual strong force, analogously to the binding of nucleons in nuclei; however, in the compact tetraquark picture, confinement is the binding force among all possible quark-(anti)quark pairs. The new quest was initiated by the discovery of $X(3872)$ in $B$ decays~\cite{Belle:2003nnu}, also denoted as $\chi_{c1}(3872)$~\cite{ParticleDataGroup:2024cfk}
after the quantum numbers were fixed to $J^{PC}=1^{++}$~\cite{LHCb:2013kgk,LHCb:2015jfc}.
The state was shortly after confirmed in various other experiments, and
a growing family of exotic hadron candidates has since been reported experimentally and investigated theoretically; see Refs.~\cite{Hosaka:2016pey,Esposito:2016noz,Guo:2017jvc,Olsen:2017bmm,Karliner:2017qhf,Kalashnikova:2018vkv,Brambilla:2019esw,Meng:2022ozq, Liu:2024uxn, Chen:2024eaq} for recent reviews. 
The $X(3872)$ continues to serve as a benchmark for testing theoretical frameworks and constraining models of hadronic structure.

Despite the large number of experimental and theoretical studies, 
several aspects of $X(3872)$ remain poorly understood. Although the $X(3872)$ mass has been measured with high precision---$3871.64(6)$~MeV according to the 2024 issue of the Review of Particle Physics~\cite{ParticleDataGroup:2024cfk}---it remains unclear whether it lies above or below the $D^0\bar{D}^{*0}$ threshold, as the uncertainty encompasses this boundary. 
Notably, the mass and width of $X(3872)$ listed in Ref.~\cite{ParticleDataGroup:2024cfk} are from averaging values extracted using the Breit-Wigner parameterization (e.g., in Refs.~\cite{BESIII:2022bse,LHCb:2020fvo,CDF:2009nxk}), which is not appropriate for describing $X(3872)$ lineshape near the $D^0\bar{D}^{*0}$ threshold, a channel to which $X(3872)$ couples strongly in $S$-wave.
While the Flatté parameterization of Ref.~\cite{Hanhart:2007yq} employed in 
Refs.~\cite{LHCb:2020xds,BESIII:2023hml} offers  improvement by including the nonanalyticity at the $D^0\bar{D}^{*0}$ threshold, it neglects the three-body effects of the $D\bar{D}\pi$ system, as discussed in Refs.~\cite{Baru:2011rs,Du:2021zzh,Ji:2022blw,Dong:2024fjk,Zhang:2024fxy} for similar systems, which could hinder a precise determination of the $X(3872)$ properties. It also has the drawback that it assumes the coupled-channel potential matrix to be non-invertible and thus might not be sufficiently general~\cite{Cohen:2004kf, Braaten:2005jj, Dong:2020hxe, Zhang:2024qkg,Sone:2024nfj,Heuser:2024biq}.
Furthermore, another potentially crucial factor is missing in all existing analyses:
In Ref.~\cite{Zhang:2024fxy}, it was shown using chiral effective field theory that data call for the existence of an isovector partner of $X(3872)$, there and here called $W_{c1}$---according to the naming scheme in Ref.~\cite{ParticleDataGroup:2024cfk} it should be $T_{c\bar c1}(3882)$ due to its quantum
numbers $I^G(J^{PC})=1^-(1^{++})$, which should not to be confused with the $T_{c\bar c1}(3900)$, also known as $Z_c(3900)$, with quantum numbers $1^+(1^{+-})$. The charged members of the same multiplet appear as virtual states in $(D\bar D^*)^\pm$ scattering. 
Their existence is supported by a recent lattice QCD calculation~\cite{Sadl:2024dbd}. The neutral $W_{c1}^0$ manifests itself as a mild cusp at the $D^+D^{*-}$ threshold whose strength is expected to be much weaker than the peak around the $D^0\bar{D}^{*0}$ threshold~\cite{Zhang:2024fxy},
which, however, distorts the lineshape and contaminates the signal of
$X(3872)$.

Therefore, in order to extract the $X(3872)$ (and $W_{c1}$) properties reliably, it is necessary to reanalyze the data in a coupled-channel framework allowing for both states and including three-body effects.
The results of this program are presented in this article.

\section{Framework}

The $X(3872)$ couples mainly to the open-charm $D^0\bar D^{*0}$ and $D^+ D^{*-}$ channels\footnote{Here and in the following, we use $D^0\bar D^{*0}$ and $D^+ D^{*-}$ to refer to the combinations with $J^{PC}=1^{++}$. The $\bar D D^*$ components are included in the calculation but not written explicitly for simplicity.} (labeled by the Greek index $\alpha{=}0,\pm$) as indicated by the large branching ratio into $D^0\bar D^{*0}$ reported in Refs.~\cite{Belle:2008fma,Li:2019kpj,Braaten:2019ags,BESIII:2020nbj,BESIII:2023hml} despite the tiny phase space---the statement remains true even with the revised branchings extracted in this work. It can also decay into several weakly coupled inelastic hidden-charm channels, such as $J/\psi\rho^0$, $J/\psi\omega$, $\chi_{cJ}\pi^0 (J=0,1,2)$, $J/\psi\gamma$, $\psi'\gamma$ and probably others. To analyze the existing data in the very limited energy region near the $D^0\bar D^{*0}$ and $D^+ D^{*-}$ thresholds {up to 3.9~GeV},
leading-order chiral effective field theory can be employed. {Higher order corrections are expected to be of the order of $(p_\text{max}/(2\mu_0))^2\approx 1\%$, where $p_\text{max}$ is the $D^0\bar D^{*0}$ center-of-mass (c.m.) momentum at 3.9~GeV and $\mu_0$ is the reduced mass.}
At this order, the $D^0\bar D^{*0}$-$D^+ D^{*-}$ coupled-channel scatterings are in $S$-waves, and the $S$-$D$ wave mixing effects do not enter. 
Because there are no common valence quark flavors in charmonia and light mesons, rescatterings within the inelastic channels can be neglected due to Okubo-Zweig-Iizuka suppression. 

The $D^\alpha\bar D^{*\alpha}$ coupled-channel scattering amplitude $T_{\alpha\beta}$ is constructed using the Lippmann-Schwinger equation (LSE), incorporating an energy-dependent $D^{*\alpha}$ decay width $\Gamma_{\alpha}(E;l)$. Here, $E$ denotes the c.m. energy relative to the $D^0\bar D^{*0}$ threshold, and $l$ represents the magnitude of the relative momentum of the intermediate $DD^*$ state. This energy-dependent width is derived from the $D^*$ self-energy arising from the $D\pi$ and $D\gamma$ channels, which consequently introduces the $D\bar D\pi(\gamma)$ three-body cuts into
the amplitude. This approach extends beyond the formalism of Refs.~\cite{Du:2021zzh,Zhang:2024fxy}, where only the imaginary part of the $D^*$ self-energy from $D\pi$ is considered and the $D\gamma$ partial width is treated as a constant.
The potential is constructed as 
\begin{align}
V(E;p^\prime,p)=V^{\mathrm{ct}}+V^{\pi}(E;p^\prime,p)+V^{\mathrm{inel}}(E)\,.
\end{align}

The $V^{\mathrm{ct}}$ term denotes a constant contact interaction with two free parameters, $C_{0X}$ and $C_{1X}$, corresponding to the isoscalar and isovector components, respectively. $V^{\pi}(E;p^\prime,p)$ represents the one-pion exchange potential. The inelastic part, $V^{\mathrm{inel}}(E)$, involves three additional free parameters, $a$, $u_{0\rho}$, and $v_0(v_1)$, and accounts for the effects of inelastic channels to which the $X(3872)$ and its isospin partner $W_{c1}^0$ can couple. More precisely, $J/\psi\rho^0$ and $J/\psi\omega$ are explicitly included since their thresholds lie very close to the energy region of interest. Other inelastic channels, such as $\chi_{cJ}\pi(\pi)$, $J/\psi \gamma$, and $\psi(2S) \gamma$, are effectively incorporated into the parameters $v_0$ (isoscalar component) and $v_1$ (isovector component). In the following analysis, we consider two extreme scenarios, Scheme-I0 $(v_1=0)$ and Scheme-I1 $(v_0=0)$, and absorb the small differences in the results into the uncertainties. Notice that these two schemes are not introduced for the purpose of comparison. Since the net effect of isospin structure of these inelastic channels is not well known, we expect that the physical situation should lie between these two extreme cases. In total, there are five free parameters in the potential: $C_{0X}$, $C_{1X}$, $a$, $u_{0\rho}$, and $v_0(v_1)$. 

\begin{figure}[h]
    \centering
    \includegraphics[width=0.85\linewidth]{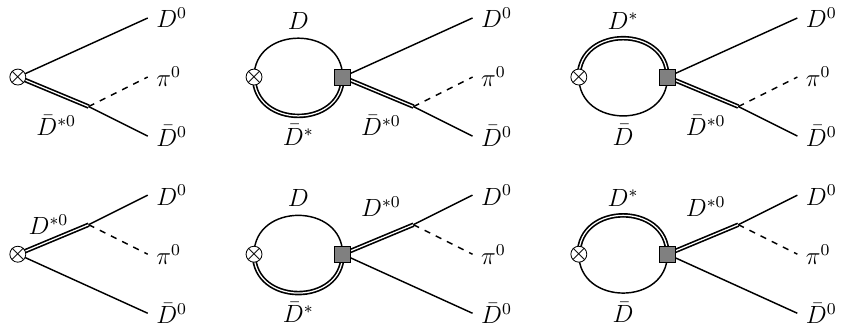}
    \caption{Diagrams for $D^0\bar D^0\pi^0$ production. The gray square represents the $D\bar D^*$ scattering in $J^{PC}=1^{++}$ channels and $\otimes$ represents the production of $D\bar D^*$ from the given $1^{++}$ source.
    }
    \label{fig:fm1}
\end{figure}
\begin{figure}[h]
    \centering
    \includegraphics[width=0.95\linewidth]{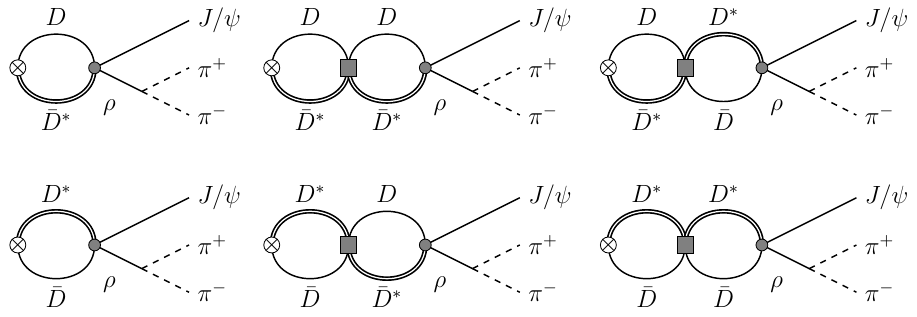}
    \caption{Similar with Fig.~\ref{fig:fm1} but for the $J/\psi\pi^+\pi^-$ production. {The gray circle stands for the elastic-inelastic transition with $\rho$-$\omega$ mixing included.}
    }
    \label{fig:fm2}
\end{figure}

Since the magnitudes of amplitudes involving inelastic-to-elastic or inelastic-to-inelastic transitions are much smaller than those of elastic channels, production via inelastic channels can be safely neglected. The production amplitudes for the $D\bar D\pi$ and $J/\psi\pi^+\pi^-$ final states, denoted by $\mathcal{A}_{D^0\bar D^0\pi^0}$ and $\mathcal{A}_{J/\psi\pi\pi}$, respectively, and illustrated in Figs.~\ref{fig:fm1} and \ref{fig:fm2}, are then constructed from the coupled-channel $DD^*$ rescattering amplitudes obtained by solving the LSE. See the Supplementary material for details of the full treatment of the coupled-channel LSE.

\section{Numerical results}

In the BESIII data~\cite{BESIII:2023hml} for the $J/\psi\pi^+\pi^-$ distribution, besides the resonance signal (taken to be from $X(3872)$ in Ref.~\cite{BESIII:2013fnz}), there is a smooth background. The same is true for the LHCb data for the same $X$ decay channel~\cite{LHCb:2020fvo}. 
We take the noninterfering backgrounds in the $J/\psi\pi^+\pi^-$ distributions from the experimental analyses and subtract them from the data to obtain the signal distributions.
In terms of the production amplitudes of two different final states,
% Eq.~\eqref{eq:Tf},
the expressions for the experimental yields read
\begin{align}
\frac{\mathrm{d\mathcal{N}_{D^0\bar D^0\pi^0}}}{\mathrm{d}E}&=\frac{1}{32\pi^3}\int_0^{p_{\rm max}}\frac{p\,{\rm d} p}{\omega_{D^0}(p)}\int_{\bar p_{\rm min}}^{\bar p_{\rm max}}\frac{\bar p\,{\rm d} \bar p}{\omega_{\bar D^0}(\bar p)}\notag\\
&\quad\times\left(|\mathcal{A}_{D^0\bar D^0\pi^0}(E,p,\bar p)|^2+f_{\rm bg}\right), \notag\\
\frac{\mathrm{d^2\,\mathcal{N}_{J/\psi\pi\pi}}}{\mathrm{d}E\,\mathrm{d}m_{2\pi}}&=\int{\rm d}\Phi_{2\pi}\frac{k_{J/\psi}(E,m_{2\pi})m_{2\pi}}{4{\pi^2} E}\notag\\
&\quad\times|\mathcal{A}_{J/\psi\pi\pi}(E,m^2_{2\pi})|^2,
\end{align}
where $p_{\rm max}$, $\bar p_{\rm min}$ and $\bar p_{\rm max}$ are determined by kinematics, $\omega_{D^0}(p){=}\sqrt{m_{D^0}^2+p^2}$ is the $D^0$ energy ($\omega_{\bar D^0}(\bar p)$ for $\bar D^0$), ${\rm d}\Phi_{2\pi}$ is the differential 2-body phase space element of $\pi^+\pi^-$ in the final state, $k_{J/\psi}(E,m_{2\pi})$ is the momentum of $J/\psi$ in the $D\bar D^*$ c.m. frame with $m_{2\pi}$ the $\pi^+\pi^-$ invariant mass. 
We have introduced a non-interfering, constant background term, denoted by $f_{\rm bg}$, to the $D^0\bar D^0\pi^0$ distribution as the background events for this process are not subtracted from the data because of the large uncertainties. 
In addition, the $P_\alpha$ parameters, which are involved in $\mathcal{A}_{D^0\bar D^0\pi^0}$ and $\mathcal{A}_{J/\psi\pi\pi}$,
parametrize the source term and at the
same time provide the normalization constants. 
Since we have two kinds of sources for producing the open-charm $D\bar D^*$ pairs, one from $e^+e^-$ annihilations and the other from $B$ decays, the production parameters denoted by $P_\alpha^{\rm B}$ and 
$P_\alpha^{\rm L}$ for the BESIII and LHCb data, respectively,  are different. 

\begin{figure*}[t]
    \centering
    \includegraphics[width=0.8\linewidth]{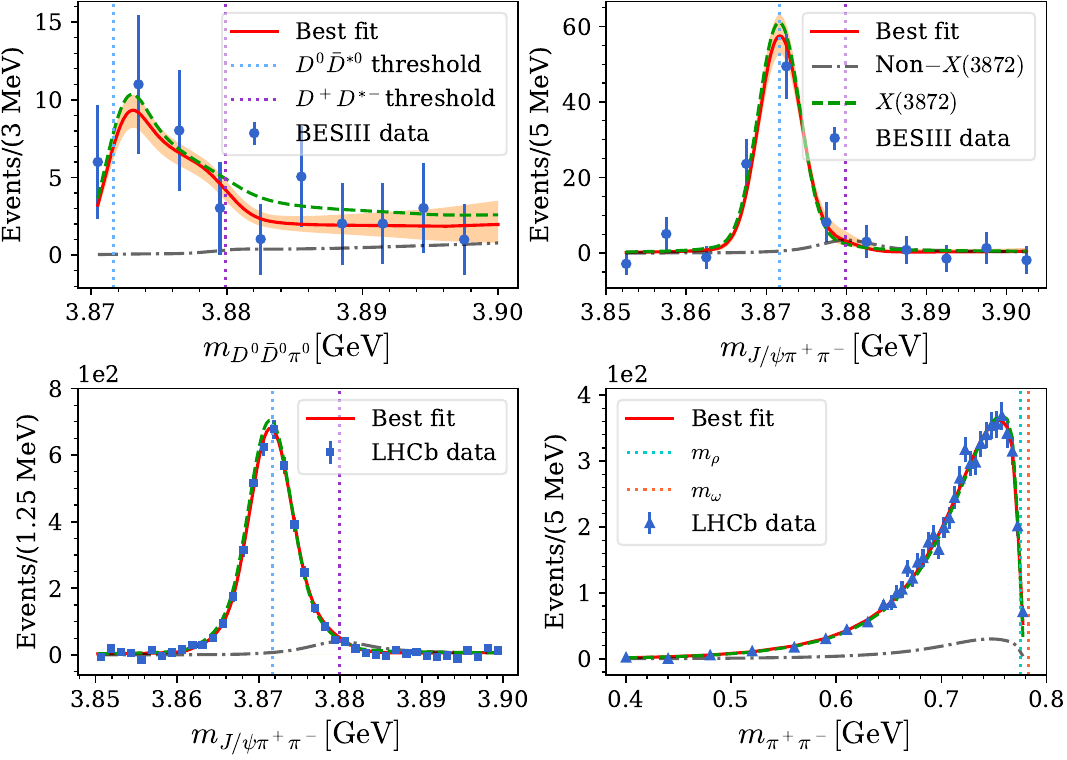}
    \caption{
Best fit in Scheme-I0 to the BESIII and LHCb data. (a) BESIII data on
$e^+e^-\to \gamma D^0\bar D^{0}\pi^0$~\cite{BESIII:2023hml}; (b) BESIII data on $e^+e^-\to \gamma J/\psi\pi^+\pi^-$~\cite{BESIII:2023hml};
(c) LHCb data on $B^+\to K^+(J/\psi\pi^+\pi^-)$, with the
$J/\psi\pi^+\pi^-$ distribution from Ref.~\cite{LHCb:2020fvo};
(d) the $\pi^+\pi^-$ distribution from Ref.~\cite{LHCb:2022jez}.
The red solid curves show the best-fit results, while the orange bands
indicate the fit uncertainty. The green dashed and gray dash-dotted curves
represent the $X(3872)$ contribution, described by a Flatt\'e formula as
detailed in the Supplementary Material, and the non-$X(3872)$ contribution,
respectively. The line shapes in Scheme-I1 are almost indistinguishable from
those in Scheme-I0.
}
    \label{fig:best_fit_all}
\end{figure*}

In total, there are 10 free parameters to be determined through simultaneously fitting to the BESIII~\cite{BESIII:2023hml} and LHCb~\cite{LHCb:2020fvo,LHCb:2022jez} by using MINUIT~\cite{James:1975dr,iminuit,iminuit.jl}:
$P_0^{\rm B}$, $P_\pm^{\rm B}$, $P_0^{\rm L}$, $P_\pm^{\rm L}$, $C_{0X}$, $C_{1X}$, $a$, $u_{0\rho}$, $v_0(v_1)$ and $f_{\rm bg}$. 
The data can be accurately described, with the best fit yielding {$\chi^2/\text{dof}= 57/86 = 0.66$ in both Scheme-I0 and Scheme-I1} for $\Lambda=1.0$~GeV, where ``dof" denotes the number of degrees of freedom in the fit. {In the following analysis, the central value of each quantity is taken as the average of the results from Scheme-I0 and Scheme-I1, while the associated uncertainty is defined as the quadrature sum of the larger of the two individual uncertainties and half the difference between the central values of the two schemes.} Fig.~\ref{fig:best_fit_all} illustrates a comparison of our results {in Scheme-I0} with the data, with the orange bands indicating the lineshapes within $1\sigma$ statistical uncertainty propagated from the experimental data.
Energy resolutions and efficiencies provided in Refs.~\cite{BESIII:2023hml,LHCb:2022jez,LHCb:2020fvo} were included in the fits.
Parameter values, the correlation matrix from the best fit and the fitting results for $\Lambda=0.6$ and $1.4$~GeV {in both Scheme-I0 and Scheme-I1} can be found in the Supplementary material.   

\begin{figure}[t]
    \centering
    \includegraphics[width=\linewidth]{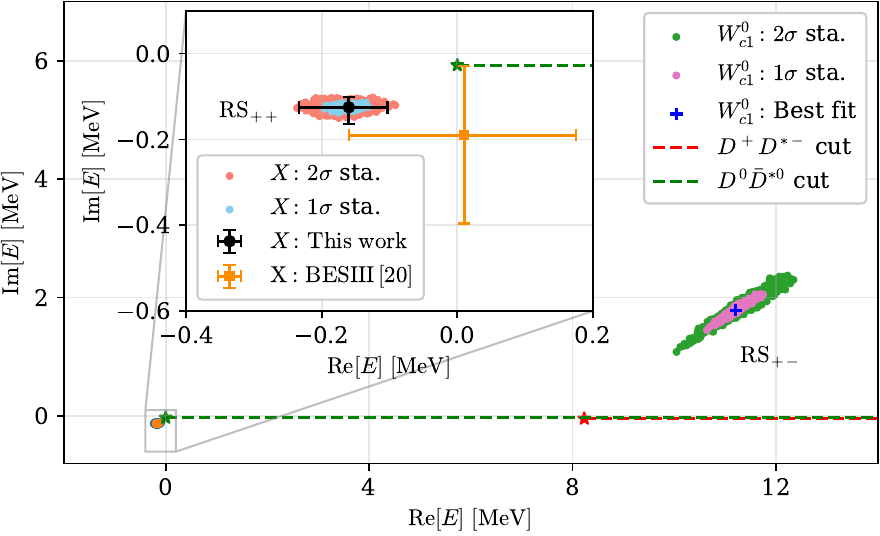}
    \caption{Pole positions of \(X(3872)\) and \(W_{c1}^0\), relative to the \(D^0\bar D^{*0}\) threshold, from our analysis in Scheme-I0 with
\(1\sigma\) and \(2\sigma\) statistical (sta.) uncertainties. Pole locations in Scheme-I1 are similar. Panel descriptions: \textbf{(a)} Full view of the pole positions of \(X(3872)\) and \(W_{c1}^0\), including the \(D^+D^{*-}\) and \(D^0\bar D^{*0}\) cuts. \textbf{(b)} Zoom-in view of the \(X(3872)\) pole region. The \(X(3872)\) pole position with total uncertainties determined in this work is compared with that from the BESIII
analysis~\cite{BESIII:2023hml}.
}
    \label{fig:polepositions}
\end{figure}

With the parameters from the best fit, the poles of the scattering amplitudes follow from solving LSE for complex energies. Labeling different Riemann sheets of the coupled-channel scattering amplitudes by the signs of the imaginary parts of momenta for the two channels $D^0\bar D^{*0}$ and $D^+D^{*-}$  (i.e., RS$_{\pm\pm}$ with RS$_{++}$ the physical sheet),
we find the $X(3872)$ pole on sheet RS$_{++}$, located at
\begin{align}
    {E_X=\left(-160^{+28+32+40}_{-38-28-60}-125^{+12+14+15}_{-13-22-28}\,i\right) \rm keV}
     \label{eq:Ex}
\end{align}
relative to the nominal $D^0\bar D^{*0}$ threshold at 3871.69~MeV, {and the real part is about $2.7\,\sigma$ smaller than zero}. 
The central value is obtained with $\Lambda{=}1.0$~GeV, the first error represents the statistical uncertainty inherited from the experimental data, the second is the systematic uncertainty from varying $\Lambda$ from 0.6 to 1.4~GeV, and the third is due to the uncertainties of $D^{(*)}$ masses and estimated by simultaneously shifting both $D^0$ and $\bar D^{*0}$ masses by {$\pm 52$} keV\footnote{{The $D^0\bar D^{*0}$ threshold is expressed as $2m_{D^0}+\Delta m$, with $\Delta m=m_{D^{*0}}-m_{D^0}$. Using the PDG uncertainties for $m_{D^0}$ and $\Delta m$~\cite{ParticleDataGroup:2024cfk}, the threshold uncertainty is calculated as $\sqrt{(2\times50)^2 + 30^2}~{\rm keV} \approx 104~{\rm keV}$.}}
 and refitting the data. 
Accordingly, the pole of $X(3872)$ is determined to be 
$\left(3871.53^{+0.06}_{-0.08}-0.13^{+0.02}_{-0.04}\,i\right)$~MeV, with errors added in quadrature. 
{The corresponding uncertainties have been added in quadrature for all other results.}
The real part is consistent with the ones obtained in the analyses performed by LHCb~\cite{LHCb:2020xds} and BESIII~\cite{BESIII:2023hml} using the generalized Flatt\'e parameterization of Ref.~\cite{Hanhart:2010wh} within 1~$\sigma$, but with a significantly reduced uncertainty {as is clearly illustrated in Fig.~\ref{fig:polepositions}}. In particular, we establish a quasi-bound state nature of $X(3872)$ with a significance of {$2.7\,\sigma$} for the first time{---the hypothesis that the pole is located at or above the $D^0\bar D^{*0}$ threshold is below 0.35\%}. {The branching ratios for the $X(3872)$ decays are shown in Table~\ref{tab:Xwidth},
based on the formalism proposed in Ref.~\cite{Heuser:2024biq}.
From the first two entries in the table the branching fraction for the decay to the $D^0\bar D^{*0}+c.c.$ channel is about 63\%.
}

\begin{table}[t]
    \caption{{Branching ratios of $X(3872)$ decays, where ``Others" represent known $\chi_{c0}\pi$, $J/\psi \gamma$ and $\psi(2S) \gamma$ channels as well as all the other unknown channels.}}
    \label{tab:Xwidth}
    \renewcommand{\arraystretch}{1.4}
\begin{tabular}{lccccc}
\toprule
Mode & $D^0\bar D^0\pi^0$ & $D^0\bar D^0\gamma$ & $J/\psi\pi^+\pi^-$ & $J/\psi\pi^+\pi^-\pi^0$ & Others \\ 
\midrule
BR (\%) & $41^{+3}_{-4}$ & $22\pm2$ & $5^{+2}_{-1}$ & $16^{+4}_{-3}$ & $16\pm2$ \\ 
\bottomrule
\end{tabular}
\end{table}

It is important to stress that there is another pole on sheet RS$_{+-}$, located at 
\begin{align}
   { \left(3.09^{+0.53+0.04+0.36}_{-0.57-0.28-0.26}+ 1.26^{+0.33+1.90+0.15}_{-0.34-0.49-0.13}\,i\right)\ \rm MeV}
   \label{eq:Wc1pole}
\end{align}
relative to the nominal $D^+ D^{*-}$ threshold at 3879.92~MeV, {where the decomposition of errors is the same as in Eq.~\eqref{eq:Ex}}. 
It corresponds to the isovector $W_{c1}^0$ state predicted recently~\cite{Zhang:2024fxy}. In contrast to the $X(3872)$ pole, this one is not directly connected to the physical region. It manifests itself as a cusp at the $D^+D^{*-}$ threshold~\cite{Zhang:2024fxy}. Both poles are presented in Fig.~\ref{fig:polepositions}.

Note that, as a result of the apparent isospin violation the two neutral states get coupled and due to this coupling the two levels repel each other, leading to an increased attraction leading to the lower state, the $X$, and a reduced attraction leading to the higher state, the $W_{c 1}^0$, which turns into a pole above threshold on RS$_{+-}$. For an extensive discussion about the impact of channel coupling on pole moving of a two-channel system, we refer to Ref.~\cite{Zhang:2024qkg}.

To quantify the impact of $W_{c1}^0$, we remove the $X(3872)$ contributions, represented by the green dashed curves from the $D^0\bar D^0\pi^0$ and $J/\psi\pi^+\pi^-$ distributions,  from the full
amplitudes, as shown by the gray dash-dotted curves in Fig.~\ref{fig:best_fit_all}. The $X(3872)$ is here described by the Flatt\'e parameterization with parameters adjusted to reproduce the $X(3872)$ pole position and residues to the elastic channels
following the recipe of Ref.~\cite{Heuser:2024biq}---for details see the Supplementary material.
The absence of a distinct structure of $W_{c1}^0$ is attributed to the dominance of $T_{00}$, which has a dip instead of a peak at the $D^+D^{*-}$ threshold~\cite{Zhang:2024fxy}, resulting from the higher production rate of the neutral channel in both reactions, i.e., $|P_0|>|P_\pm|$---the best fit values are {$P_{ \pm}^{\mathrm{B}} / P_0^{\mathrm{B}}=0.6(4)$ and $P_{ \pm}^{\mathrm{L}} / P_0^{\mathrm{L}}=0.5(1)$}. The reason for this pattern is that in $e^+e^-$ annihilations, $X(3872)$ is produced dominantly through the radiative decay of $Y(4230)$~\cite{Guo:2013zbw, BESIII:2013fnz,BESIII:2019qvy}, which has large coupling to $D\bar D_1$ in $S$-wave~\cite{Wang:2013cya,vonDetten:2024eie} and the radiative decay of $D_1^0\to D^{*0}\gamma$ is significantly larger than that of $D_1^+\to D^{*+}\gamma$~\cite{Korner:1992pz, Fayyazuddin:1994qu}. For the $B^+$ decays, one finds experimentally $\operatorname{Br}(B^+\to D^0\bar D^{*0}K^+)> \operatorname{Br}(B^+\to D^+D^{*-}K^+)$~\cite{BaBar:2003qrf,BaBar:2010tqo}. 
{We notice that the dip in $|T_{00}|$ at the $D^+D^{*-}$ threshold is a universality feature for  $D^+D^{*-}\to D^+D^{*-}$ having an unnaturally large scattering length, as shown in Ref.~\cite{Dong:2020hxe}.}

\begin{figure*}[tb]
    \centering
    \includegraphics[width=0.7\linewidth]{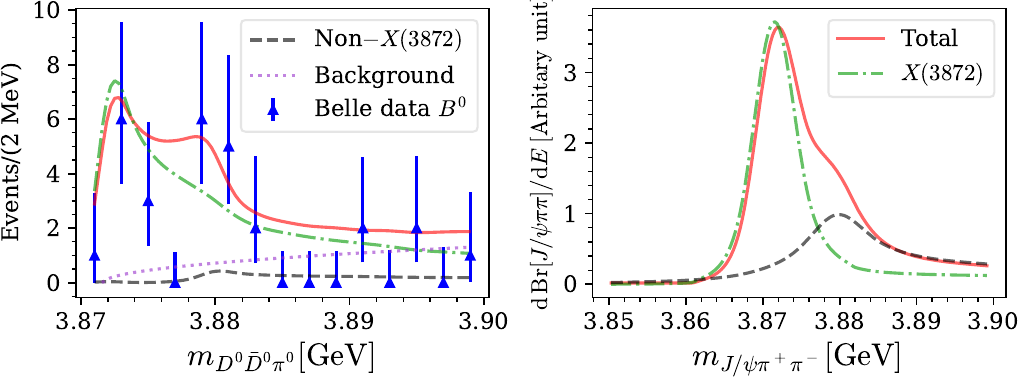}
    \caption{Predicted line shapes for $B^0$ decays using $P_\pm/P_0=2$ in Scheme-I0. (a) The $D^0\bar D^0\pi^0$ invariant-mass distribution in comparison with the Belle data for
\(B^0\to K^0D^0\bar D^0\pi^0\)~\cite{Belle:2023zxm}; (b) the $J/\psi\pi^+\pi^-$ invariant-mass distribution. 
Convolutions with the Belle~\cite{Belle:2023zxm} and
LHCb~\cite{LHCb:2020fvo} energy resolutions are considered in (a) and (b), respectively.
}
    \label{fig:predictedls}
\end{figure*}

In contrast, in  $B^0$ decays the branching ratio to $K^0D^+ D^{*-}$ is six times that of $K^0D^0\bar D^{*0}$~\cite{BaBar:2010tqo}. Therefore, we expect $|P_{\pm}|>|P_0|$ for the analogous $B^0$ decays, and thus the signal of $W_{c1}^0$ in $B^0$ decays should be more pronounced than that in the $B^+$ decays. Indeed, by fixing $P_{\pm}/P_0=2$, reasonable for $B^0$ decays, the predicted lineshapes exhibit this feature, as shown in Fig.~\ref{fig:predictedls}. For comparison we also show the Belle data  for $B^0\to K^0D^0\bar D^{0}\pi^0$~\cite{Belle:2023zxm}, which
are indeed consistent with
a sizable $W_{c1}^0$ contribution, although current data quality does not allow for a firm conclusion. A similar structure is anticipated in the $D^0\bar D^0\gamma$ distribution for $B^0\to K^0D^0\bar D^{0}\gamma$. We expect that the $W_{c1}^0$-induced peak can be unambiguously identified with the full Belle~II statistics. The predicted nontrivial right shoulder of the $J/\psi\pi^+\pi^-$ distribution in the right panel of Fig.~\ref{fig:predictedls} can be checked through $B^0\to K^0J/\psi\pi^+\pi^- $ at both LHCb and Belle~II. Furthermore, one can also detect the charged partners of $W_{c1}^0$ in the charged channels, where a threshold cusp should appear at the $D^+\bar D^{*0}$ threshold, as discussed in Ref.~\cite{Zhang:2024fxy}. However,
the neutral channel has the advantage that the signal gets enhanced by the interference with the signal
from $X(3872)$. 

We also notice the conflict of more than $5\sigma$ for $\Gamma_{X(3872)\to J/\psi\gamma}/\Gamma_{X(3872)\to \psi(2S)\gamma}$ between the measurements by LHCb in $B^{+}\to K^+ X(3872)$~\cite{LHCb:2024tpv} and by BESIII in $e^+e^-\to \gamma X(3872)$~\cite{BESIII:2020nbj}. This conflict may be resolved by considering that the observed signals contain different $W_{c1}$ contributions in these two cases, arising from the different production rates of $D^0\bar D^{*0}$ and $D^+D^{*-}$ in $B^+$ decays and $e^+e^-$ annihilations.

Another important property related to the nature of $X(3872)$ is the significant isospin breaking in its decays, quantified by the ratio of its couplings to $J/\psi\rho^0$ and $J/\psi\omega$, defined as $R_X \equiv |g_{XJ/\psi\rho}/g_{XJ/\psi\omega}|$~\cite{Suzuki:2005ha,Gamermann:2009uq,Hidalgo-Duque:2012rqv}  with more details in the Supplementary material. 
While there have been determinations of $R_X$~\cite{LHCb:2022jez,Hanhart:2011tn,Wang:2022vjm,Dias:2024zfh}, none of them so far has considered that the data contain $W_{c1}^0$  contributions.
In the formalism presented here, the isospin breaking parameter $R_X$ is given by the ratio of the production amplitudes of $J/\psi\rho$ and $J/\psi\omega$ from a given source at the $X(3872)$ mass. Explicitly, {evaluated at the pole,} we have 
% \begin{align}
%     \hspace{-3mm} R_X {=}\Bigg|\frac{\int_0^\Lambda {{l^2{\rm d}{l}}}\, U_\alpha(E_X,l)G_{\alpha\beta}(E_X,l)u_{\beta \rho}}{\int_0^\Lambda {{l^2{\rm d}{l}}}\, U_\alpha(E_X,l)G_{\alpha\beta}(E_X,l)u_{\beta \omega}}\Bigg|={0.26(2)} , \label{eq:RX}
% \end{align} 
\begin{align}
     R_X={0.26(2)} , \label{eq:RX}
\end{align} 
which is {consistent with} the value $0.26(3)$ obtained without the $W_{c1}^0$ contribution~\cite{Dias:2024zfh}.

The compositeness of $X(3872)$ can be calculated using the range-corrected formula~\cite{Li:2021cue} as
\begin{align}
    X=1-\exp \left(\frac{1}{\pi} \int_0^{
    \frac{\Lambda^2}{2\mu_0}}\!\! d E \frac{\operatorname{Re}\delta(E)}{E-\operatorname{Re}E_X}\right)  = {0.97(2)},
\end{align}
where $\delta(E)$ is the $S$-wave $D^0\bar D^{*0}$ phase shift with the convention that it vanishes at the threshold,  and the upper bound of the integration is fixed by the cutoff in our calculation. The phase shift $\delta(E)$ is determined by $k\cot \delta=-\frac{2\pi}{\mu_0 T_{00}}+ik$ with $k$ the on-shell momentum of the $D^0 \bar D^{*0}$ state. Here the effect of $D^*$ decays was removed from the expression by using the real part of the phase only. The uncertainty is dominated by $\mathcal{O}(\gamma_X^2/\Lambda^2)$ corrections~\cite{Li:2021cue}, with $\gamma_X$ the $X(3872)$ binding momentum. It is consistent with the $X(3872)$ being a molecular state, and with the value {$0.97^{+0.03}_{-0.07}$} determined using scattering length and effective range with the formula of Ref.~\cite{Matuschek:2020gqe}. Expanding $T_{00}$ around the $D^0\bar D^{*0}$ complex threshold~\cite{Braaten:2009jke, Baru:2021ldu} by $T_{00}=-\frac{2\pi}{\mu_0}\,\left(\frac{1}{a_0}+\frac{r_0}{2} k^2-ik \right)^{-1}$ we find the scattering length 
{$a_0=\left(-9.6^{+1.5}_{-1.7}+2.8^{+1.2}_{-0.8}\,i\right)\ \rm fm$} and effective range  
{$r_0'=-0.3^{+0.3}_{-0.9}\ \rm fm$}, which is the one defined in Ref.~\cite{Baru:2021ldu} after correcting for channel coupling effects. For completeness, we have also performed the same expansion for the $T_{\pm\pm}$ channel around $D^{+}D^{*-}$ threshold. This yields the corresponding scattering length $a_{\pm}=3.3\pm1.2 +  (0.1\pm0.7) i \ \rm fm$ and effective range $r_{c}=-0.4\pm0.8+(1.5\pm0.5) i \ \rm fm$. Compared with the results of Ref.~\cite{Song:2023pdq}, we find that the central value magnitude of the scattering length in the neutral channel is smaller in our case, which can be traced back to the deeper binding energy of the $X(3872)$. For the charged channel, our scattering length is
significantly larger than that of 
Ref.~\cite{Song:2023pdq} reflecting the presence
of the near threshold virtual state, $W_{c1}^0$.
Ref.~\cite{Song:2023pdq} did not include isovector interactions and accordingly the scattering length
for the charged channel reported in that work is much smaller.

\section{Conclusion}
We have determined the $X(3872)$ properties with unprecedented accuracy by fitting a large set of
experimental data with an improved formalism. More precisely, 
we demonstrate that the data are consistent with $X(3872)$ emerging as a molecular state from $D^0 \bar D^{* 0}$ and $D^+ D^{*-}$ coupled-channel interactions. The inelastic coupled channels $J/\psi \rho$ and $J/\psi \omega$ are taken into account explicitly. The $X(3872)$ is determined to be a (quasi-)bound state with a significance of {$2.7\,\sigma$}, with a pole located at {$\left(-160^{+57}_{-74}-125^{+23}_{-38}\,i\right) \rm keV$} relative to the $D^0 \bar D^{*0}$ threshold. 
{Accordingly, the $X(3872)$ pole is determined to be at
$\left(3871.53_{-0.08}^{+0.06}-0.13_{-0.04}^{+0.02}\,i\right)$~MeV. The precision significantly improves over the result reported by BESIII~\cite{BESIII:2023hml}, see the comparison in Fig.~\ref{fig:polepositions}.}
The isospin breaking ratio of its decays into $J/\psi\rho^0$ and $J/\psi\omega$ is determined as {$R_X{=}0.26(2)$}. 

We have also determined the pole of the isospin-vector partner of $X(3872)$, $W_{c1}^0$ predicted in~\cite{Zhang:2024fxy}, to be at {$\left(3.1\pm0.7+ 1.3^{+1.9}_{-0.6}\,i\right)$}~MeV relative to the $D^+ D^{*-}$ threshold on an unphysical Riemann sheet {with errors in Eq.~\eqref{eq:Wc1pole} added in quadrature}. 
Its significance is found to be very high by comparing an alternative fit scheme without the $W_{c1}^0$ contribution; see the Supplementary material. 

The $W_{c1}^0$ leads to very mild distortions of the $D^0\bar D^0\pi^0$ and $J/\psi\pi^+\pi^-$ distributions from the $X(3872)$ line shapes in both $e^+e^-$ collisions and $B^+$ decays.
Its signal should be more clearly visible in reactions where $D^+D^{*-}$ is more frequently produced than $D^0\bar D^{*0}$, such as $B^0\to K^0 D\bar D^*$ decays, which can be verified at Belle II and LHCb. {In particular, the nontrivial prediction of a right shoulder around 3.88~GeV in the $J/\psi\pi^+\pi^-$ distribution in $B^0\to K^0J/\psi\pi^+\pi^- $ is ready to be tested as the $J/\psi\pi^+\pi^-$ final state can be easily reconstructed.}

Note that the existing data can also be described by a single Flatt\'e form omitting the $W_{c1}$ similar to what was used in the experimental analyses~\cite{LHCb:2020xds,BESIII:2023hml}. However, we found using the compositeness criteria that such kind of Flatt\'e also describes a molecular state, which is way more consistently treated with the formalism presented here---a statement that can be tested experimentally in $B^0$ decays!

With the confirmation of $W_{c1}$, an SU(3) flavor multiplet structure for hidden-charm hadronic molecules is emerging. Unlike compact tetraquark models, where all states are bound states of quarks and antiquarks, the molecular picture allows for a richer spectrum of states, including both bound and virtual states, with the latter showing up as threshold cusps.
The $X(3872)$ and $W_{c1}$ are prime examples of such a spectrum. The situation reminds of the two-nucleon systems, where the deuteron is an isoscalar proton-neutron bound state and a virtual pole exists in the isovector sector.
More similar structures are expected to be found in the future. Investigating these states will provide deeper insights not only into the hadron spectrum but also fundamental questions, such as why thousands of atomic nuclei exist while no evidence for compact multiquark states with $3n$ ($n\geq2$) quarks has been found, ultimately shedding light on the nature of the inner workings of the strong force.


\bigskip
\begin{acknowledgments}
We are grateful to Vanya Belyaev, Ji-Bo He, Xiao-Yu Li, Tomasz Skwarnicki, Chang-Zheng Yuan and Zhen-Hua Zhang for fruitful discussions and to Eulogio Oset for his careful reading of this manuscript and important comments. This work is supported in part by the National Key R\&D Program of China under Grant No. 2023YFA1606703; by the National Natural Science Foundation of China (NSFC) under Grants No. 12125507, No. 12361141819, and No. 12447101;  by the Chinese Academy of Sciences (CAS) under Grants No.~YSBR-101, and by the Deutsche Forschungsgemeinschaft (DFG, German Research Foundation) under Germany's Excellence Strategy – EXC 3107 – Project-ID~533766364. In addition, U.-G.M. and C.H. thank the CAS President's International Fellowship Initiative (PIFI) under Grants No. 2025PD0022 and No. 2025PD0087,
respectively, for partial support. 
\end{acknowledgments}

\bibliography{refs}

\begin{onecolumngrid}
\begin{appendix}
\section{Lippmann-Schwinger equation and the $D^*$ self-energy}\label{app:Dstarselfenergy}

The $D^0\bar D^{*0}-D^+D^{*-}$ coupled-channel scattering amplitude is constructed by the Lippmann-Schwinger equation (LSE),
\begin{align}
    T_{\alpha\beta}&(E;p^\prime,p)=\, V_{\alpha\beta}(E;p^\prime,p)  +\sum_{\mu\nu}\int_0^\Lambda\frac{l^2dl}{2\pi^2}V_{\alpha\mu}(E;p',l)G_{\mu\nu}(E;{l})T_{\nu\beta}(E;l,p), \label{eq:LSE}
\end{align}
where $E$ is the c.m. energy relative to the $D^0\bar D^{*0}$ threshold, $p$ $(p')$ is the magnitude of the incoming (outgoing) momentum, and $\Lambda$ is a hard cutoff. 
$G$ is the diagonal matrix of two-body propagators,
\begin{align}
    G_{\alpha\beta}(E;l)=\frac{\delta_{\alpha\beta}}{E-\Delta_{\alpha0}-\frac{l^2}{2\mu_{\alpha}}+\frac i2{\Gamma_\alpha(E;l)}}, \label{eq:G}
\end{align}
with $\Delta_{\alpha0}$ the threshold difference between channel-$\alpha$ and channel-0 and $\mu_{\alpha}$ the reduced mass of particles in channel-$\alpha$.

The $D^*$ propagator is given by
\begin{align}
   \frac{1}{w^2-m_0^2-\sum\limits_{i=\pi,\gamma} g_{i}^2 k^2_{i}(w)\mathcal{S}_{i}(w)}=\frac{1}{w^2-m_{D^*}^2+im_{D^*}\Gamma_{D^*}(w)},
\end{align}
where $w$ denotes the $D^*$ invariant mass, $m_0$ is the bare mass, and $m_{D^*}$ corresponds to its physical (renormalized) mass. The quantity $k_{i}$ denotes the momentum of $D$ in the rest frame of the decaying $D^*$, for the decay channel $D^* \to D+i$, with $i=\pi,\gamma$. The coupling constant $g_{i}$ characterizes the strength of the $D^* \to D+i$ vertex, which can be determined by the corresponding physical partial width of $\Gamma_{D^*\to D+i}^{\rm phy}$. Explicitly, we have
\begin{align}
  \Gamma_{D^*\to D+i}^{\rm phy}=\frac{1}{2 w} g_{i}^2k^2_{i}(w)\left.\frac{k_{i}(w)}{4\pi w}\right|_{w=m_{D^*}}.
\end{align} 
The corresponding contribution of the $D+i$ loop to the $D^*$ self-energy is given by:
\begin{align}
 \mathcal{S}_{i}(w)=&\,\frac{k_{i}(w)}{16\pi  w}\left[\log(w^2-\Delta_i+2wk_{i}(w))+\log(w^2+\Delta_i+2wk_{i}(w))\right.\notag\\
    &\left.-\log(-w^2+\Delta_i+2wk_{i}(w))-\log(-w^2-\Delta_i+2wk_{i}(w))\right],
\end{align}
where $\Delta_i=m_D^2-m_i^2$, with $m_D$ and $m_i$ denoting the masses of the $D$ meson and particle $i = \pi$ or $\gamma$, respectively. The branch cuts for the square root and logarithmic functions are taken along the negative imaginary axis to ensure the continuity between the upper edge of the first Riemann sheet of the complex $w$ plane and the lower edge of the second Riemann sheet. 
The bare mass $m_0^2$ can be fixed by the renormalization condition,
\begin{align}
    \left(m_{D^*}-\frac{i}{2}\Gamma_{D^*}^{\rm phy}\right)^2-m_0^2-\sum_{i=\pi,\gamma} g_{i}^2k^2_{i}(m_{D^*})\mathcal{S}_{i}(m_{D^*})=0,
\end{align}
where $\Gamma_{D^*}^{\rm phy}$ is the total physical decay width of the $D^*$. Consequently, the energy-dependent total  width of the $D^*$ can be expressed as
\begin{align}
    \Gamma_{D^*}(w)=\frac{i}{m_{D^*}}\left(m_0^2+\sum\limits_{i=\pi,\gamma}g_{i}^2 k^2_{i}(w)\mathcal{S}_{i}(w)-m_{D^*}^2\right),
\end{align}
which admits an analytic continuation to the below-threshold region.
In the main text, the quantity $\Gamma_\alpha$ corresponds to the energy-dependent width of the intermediate $D^{*\alpha}$ meson and is defined as 
\begin{align}
    \Gamma_\alpha(E; l) = \left.\Gamma_{D^{*\alpha}}(w) \right|_{w= E -\Delta_{\alpha0}+ m_{D^{*\alpha}} - \frac{l^2}{2\mu_\alpha}},
\end{align}
where $E$ is the energy relative to the $D^0 \bar D^{*0}$ threshold, $l$ denotes the magnitude of the $D$ or $D^*$ momentum in the $D D^*$ c.m. frame, and $\mu_\alpha$ is the reduced mass of channel $\alpha$.

\section{Technical details for the potential}\label{app:potential}

$V^{\mathrm{ct}}$ of the Eq.(1) of the main text, is the constant contact interaction, which in channel space reads~\cite{Hidalgo-Duque:2012rqv}
\begin{align}  
V^{\mathrm{ct}}=\frac{1}{2}
\begin{pmatrix} C_{0X}{+}C_{1X} & C_{0X}{-}C_{1X} \\ C_{0X} {-}C_{1X} & C_{0X} {+}C_{1X} \\
\end{pmatrix},
\end{align}
with $C_{0X}$ and $C_{1X}$ the isoscalar and isovector low-energy constants, respectively. 
The one-pion-exchange potential, $V^{\pi}$, introduces an additional three-body cut into the calculation necessary for theoretical consistency of the formalism~\cite{Baru:2011rs,Du:2021zzh,Zhang:2024fxy}. $V_{\mathrm{inel}}$ accounts for effects of the inelastic channels that $X(3872)$ and its isovector 
partner $W_{c1}^0$ can couple to.
This can be done in a way consistent with unitarity~\cite{Hanhart:2015cua, Guo:2016bjq}, which gives
\begin{align}
  \hspace{-2.5mm}  V^{\mathrm{inel}}_{\alpha\beta}(E)&= - i  {f_{\alpha\beta}^{\rm NR}}\sum_{j=\rho,\omega}\int v_{\alpha j}(s){{\rho}_j(E,s)}v^*_{\beta j}(s) {\varrho}_{j}(s)\,{\rm d}s  -\frac{i}{2}
\begin{pmatrix} v_{0}{+}v_{1} & v_{0}{-}v_{1} \\ v_{0}{-}v_{1} & v_{0}{+}v_{1} \\
\end{pmatrix}_{\alpha\beta}, \label{eq:inel}
\end{align}
{where the first term represents the contributions from the channels $J/\psi\rho,J/\psi\omega$, labeled by the Latin indices $j=\rho,\omega$, while the second term stands for collective contributions from all other possible inelastic channels. $f_{\alpha\beta}^{\rm NR}=1/{\sqrt{16m_{\alpha1}m_{\alpha_2}m_{\beta1}m_{\beta_2}}}$ is introduced for the nonrelativistic normalization in Eq.~\eqref{eq:G} with $m_{\alpha n}$ the mass of the $n$-th particle in channel $\alpha$.} 
The phase space factor is ${\rho_j(E,s){=}{q_{j{\rm cm}}(E,s)}/{(8\pi E)}}$, with $q_{j{\rm cm}}$ the magnitude of c.m. momentum in channel-$j$ and $\sqrt s$ the invariant mass of the light quark system. 
The vector-meson spectral function,
\begin{equation}
    \varrho_j(s)= -N_{j}\, {\rm Im}[G_j(s)],
\end{equation}
accounts for the finite width effects of the $\rho,\omega$ mesons, where $N_{j}$ is fixed via the normalization $\int ds \varrho_j(s)=1$. We take the $\omega$ propagator of the Breit-Wigner type, $G_\omega(s)=(s-m_\omega^2+i m_\omega\Gamma_\omega)^{-1}$ with $\Gamma_\omega=8.68$~MeV~\cite{ParticleDataGroup:2024cfk}, {which is legitimate because of the small width.} For $G_\rho$, we use the best available spectral function in terms of the Omnès function~\cite{Omnes:1958hv}, as done in Ref.~\cite{Dias:2024zfh}. {It 
contains the full information on $\pi\pi$ $P$-wave scattering 
and not just the $\rho$ resonance. 
} 
The transition of elastic channel-$\alpha$ to inelastic channel-$j$ is parameterized as
\begin{align}
    v_{\alpha j}(s)=(1+a s)u_{\alpha k}
\begin{pmatrix}
1 & \epsilon_{\rho\omega}G_\rho(s) \\ \epsilon_{\rho\omega} G_\omega(s) & 1 \\
\end{pmatrix}_{kj}. \label{eq:vinel}
\end{align}
{It considers the isospin breaking from $\rho$-$\omega$ mixing, characterized by the mixing angle $\epsilon_{\rho\omega}{=}3.35(8){\times}10^{-3}~{\rm GeV}^2$~\cite{Dias:2024zfh},  which is, however, enhanced by a factor of $m_\omega/\Gamma_\omega\approx 90$.}
Here $u_{\alpha k}$ are bare couplings. The renormalized ones, due to isospin symmetry{---it holds at the percent level for the couplings---} satisfy 
\begin{equation}
u_{0\rho}^R = u_{0\omega}^R = -u_{\pm\rho}^R = u_{\pm\omega}^R \ ,
\label{eq:udef}
\end{equation}
where $u^R_{\alpha k}=\mathcal{G}_{\alpha\alpha} u_{\alpha k}$ with $u_{\pm\rho}$ a free parameter to be fitted and $\mathcal{G}_{\alpha\alpha}=\int_0^\Lambda\frac{l^2dl^2}{2\pi^2}G_{\alpha\alpha}(0;l)$.
We have introduced a first order polynomial in $s$ as in Refs.~\cite{Stollenwerk:2011zz,Hanhart:2013vba}, 
{which is shared by the two channels because the $\rho$ and $\omega$ are in the same flavor multiplet.} 
The slope $a$ is a parameter to be determined from the fit. 
For inelastic channels other than $J/\psi\rho^0$ and $J/\psi\omega$, {such as $\chi_{cJ}\pi(\pi), J/\psi \gamma$ and $\psi(2S) \gamma$,} the phase space factors are approximately constants in the narrow energy region of interest, and they also contribute to the imaginary part of the potential. However, due to limited statistics and poor energy resolution in current experimental data for those channels~\cite{BESIII:2019esk,LHCb:2024tpv,BESIII:2020nbj,Belle:2011wdj,BaBar:2008flx}, it is challenging to determine their individual contributions reliably. Therefore, we introduce two parameters, $v_0$ and $v_1$ in the second term in Eq.~\eqref{eq:inel}, to collectively represent the effects of these inelastic channels with isospin-0 and isospin-1, respectively. 
\section{ Production amplitudes of $D^0\bar D^0\pi^0$ and $J/\psi\pi^+\pi^-$}\label{app:prodamp}
 We adopt the following charge conjugation convention:
\begin{align}
    \mathcal C\ket{D}=\ket{\bar D},\quad \mathcal C\ket{D^*}=-\ket{\bar D^*}
\end{align}
with $\mathcal C$ the charge conjugation operator. Under this convention, the wave function for $J^{PC}=1^{++}$ states read
\begin{align}
    \ket{1^{++}}=\frac{1}{\sqrt{2}}\left(\ket{D\bar D^*}-\ket{\bar DD^*}\right).
\end{align}
More precisely, the $D^0 \bar D^{*0}$ and $D^+D^{*-}$ mentioned in the main text represents:
\begin{align}
    \ket{1^{++}}_0&=\frac{1}{\sqrt{2}}\left(\ket{D^0\bar D^{*0}}-\ket{\bar D^0 D^{*0}}\right) \notag\\
    \ket{1^{++}}_\pm&=\frac{1}{\sqrt{2}}\left(\ket{D^+\bar D^{*-}}-\ket{\bar D^- D^{*+}}\right)
\end{align}

With the amplitude solved from LSE, the production of channel-$\alpha$ from a given $J^{PC}=1^{++}$ source can be constructed as
\begin{align}
    U_\alpha(E,p)=  P_\alpha+\int_0^\Lambda \frac{{l^2{\rm d}{l}}}{2\pi^2}P_\mu G_{\mu\nu}(E,l)T_{\nu\alpha}(E;l,p),
    \label{eq:production}
\end{align}
where $P_\alpha$ is the direct production amplitude of the neutral or charged $D\bar D^*$ channel. {The production via inelastic channels can be safely neglected, since the amplitudes involving inelastic-to-elastic or inelastic-to-inelastic transitions are much smaller than those of elastic channels.} Then the production amplitudes of $D^0\bar D^0\pi^0$ and $J/\psi\pi^+\pi^-$ read
\begin{align}
    &\mathcal{A}_{D^0\bar D^0\pi^0}(E,p,\bar p)=U_\alpha(E,p)F_{\alpha}(p)+U_\alpha(E,\bar p)F_{\alpha}(\bar p), \notag\\
    &\mathcal{A}_{J/\psi\pi\pi}(E,s)=\int_0^\Lambda \frac{{l^2{\rm d}{l}}}{2\pi^2} U_\alpha(E,l){G^{\rm R}_{\alpha\beta}(E,l)}H_\beta(s),\label{eq:Tf}
\end{align} 
where {$G^{\rm R}_{\alpha\beta}(E,l)=f_{\alpha\beta}^{\rm NR}G_{\alpha\beta}(E,l)$}, $p$ and $\bar p$ are the momenta of the final state $D^0$ and $\bar D^0$ in the $D\bar D^*$ c.m. frame, respectively.
$F(p)$ and $H(s)$ represent the transition of $D\bar D^*$ to $D^0\bar D^0\pi^0$ and $J/\psi\pi^+\pi^-$, respectively,
\begin{align}
     F_{\alpha}(p)&=\delta_{\alpha 0}g_{D^*D\pi} G_{D^{*0}}(p)q_{1\pi}(p), \notag\\
     H_\alpha(s)&= \sqrt{2}g_{\rho\pi\pi} v_{\alpha \rho} G_{\rho}(s) q_{2\pi}(s), \label{eq:FandH}
\end{align}
where $q_{1\pi}$ ($q_{2\pi}$) is the momentum of $\pi^0$ ($\pi^+$) in the rest frame of $D^*$  ($\rho^0$), $G_{D^{*0}}(p)\equiv G_{00}(E;p)/(2m_{D^{*0}})$ is the $D^{*0}$ propagator, and $g_{\rho\pi\pi}$ and $g_{D^*D\pi}$ are the coupling constants for the $\rho\pi\pi$ and $D^*D\pi$ vertices, respectively.

\section{Evaluation of isospin breaking}
 The isospin breaking of $X(3872)$ is defined by
 \begin{align}
    R_X=\left|\frac{g_{XJ/\psi\rho}}{g_{XJ/\psi\omega}}\right|=\left|\frac{\mathcal A_{X \to J/\psi \rho}}{\mathcal A_{X \to J/\psi \omega}}\right|
    \end{align}
   which is obtained by evaluating the ratio of the production amplitude of $J/\psi\rho$ and $J/\psi\omega$ at the pole position of $X(3872)$,
   \begin{align}
   R_X=\left|\frac{\int_0^\Lambda {{l^2{\rm d}{l}}}\, U_\alpha(E_X,l)G_{\alpha\beta}(E_X,l)u_{\beta \rho}}{\int_0^\Lambda {{l^2{\rm d}{l}}}\, U_\alpha(E_X,l)G_{\alpha\beta}(E_X,l)u_{\beta \omega}}\right|=0.26(2).
   \end{align}
   To get more insights of this quantity, we can apply the onshell approximation to get
    \begin{align}
R_X=\left|\frac{g_{X,0}\mathcal G_{00}u_{0\rho}+g_{X,\pm}\mathcal G_{\pm\pm}u_{\pm\rho}}{g_{X,0}\mathcal G_{00}u_{0\omega}+g_{X,\pm}\mathcal G_{\pm\pm}u_{\pm\omega}}\right|=\left|\frac{g_{X,0}u^{\rm R}_{0\rho}+g_{X,\pm}u^{\rm R}_{\pm\rho}}{g_{X,0}u^{\rm R}_{0\omega}+g_{X,\pm}u^{\rm R}_{\pm\omega}}\right|,
\end{align}
where $g_{X,\alpha}$ represents the coupling of $X(3872)$ to channel $\alpha$, given in Eq.~\eqref{eq:gs} below.
Given the relation in Eq.~\eqref{eq:udef}, the isospin breaking is reduced to the ratio in terms of physical couplings, 
\begin{align}
    R_X=\left|\frac{g_{X,0}-g_{X,\pm}}{g_{X,0}+g_{X,\pm}}\right|,
\end{align}
and thus cutoff-independent. This is
in line with the relation used in Refs.~\cite{Li:2012cs,Meng:2021kmi,Zhang:2024fxy}.

\section{Parameter values and lineshapes from the best fits}\label{app:params}

The parameter values and the correlation matrix from the best fits { in Scheme-I0 and Scheme-I1 are listed in Tables~\ref{tab:paras-I0} and~\ref{tab:paras-I1}, respectively.} We show the lineshapes of the best fits with different values for the cutoff parameter $\Lambda$ in Fig.~\ref{fig:difflamb}. The best fits have similar quality and the difference in lineshapes between different $\Lambda$ values is almost invisible as it should be the case for a properly renormalized effective field theory.

\begin{table}[htbp]
\caption{The parameter values for $\Lambda=0.6$, $1.0$, and $1.35~\mathrm{GeV}$ and the correlation matrix for $\Lambda=1.0~\mathrm{GeV}$ from the best fit in Scheme-I0. }
\label{tab:paras-I0}
\renewcommand{\arraystretch}{1.5}
\begin{tabular}{lcccrrrrrrrrrr}
\toprule
Parameters & $\Lambda=0.6~\mathrm{GeV}$ & $\Lambda=1.0~\mathrm{GeV}$ & $\Lambda=1.35~\mathrm{GeV}$ & \multicolumn{10}{c}{Correlation matrix ($\Lambda=1.0~\mathrm{GeV}$)} \\
\midrule
$P_0^{\rm B}/10^3~[\mathrm{GeV}^{-3/2}]$ & $4.17\pm1.40$ & $3.22\pm0.79$ & $2.21\pm0.60$ & $1.00$ &&&&&&&&& \\
$P_{\pm}^{\rm B}/P_0^{\rm B}$ & $0.61\pm0.29$ & $0.59\pm0.32$ & $0.63\pm0.46$ & \cellcolor{negcolor!74}$-0.74$ & $1.00$ &&&&&&&& \\
$P_0^{\rm L}/10^3~[\mathrm{GeV}^{-3/2}]$ & $7.55\pm2.30$ & $5.86\pm1.14$ & $4.01\pm0.69$ & \cellcolor{poscolor!81}$0.81$ & \cellcolor{negcolor!30}$-0.30$ & $1.00$ &&&&&&& \\
$P_{\pm}^{\rm L}/P_0^{\rm L}$ & $0.53\pm0.09$ & $0.542\pm0.048$ & $0.63\pm0.08$ & \cellcolor{negcolor!16}$-0.16$ & \cellcolor{poscolor!12}$0.12$ & \cellcolor{negcolor!25}$-0.25$ & $1.00$ &&&&&& \\
$C_{0X}~[\mathrm{GeV}^{-2}]$ & $-12.52\pm1.38$ & $-3.60\pm0.12$ & $0.251\pm0.043$ & \cellcolor{poscolor!58}$0.58$ & \cellcolor{negcolor!28}$-0.28$ & \cellcolor{poscolor!72}$0.72$ & \cellcolor{negcolor!51}$-0.51$ & $1.00$ &&&&& \\
$C_{1X}~[\mathrm{GeV}^{-2}]$ & $-17.28\pm0.94$ & $-11.22\pm0.30$ & $-8.26\pm0.15$ & \cellcolor{poscolor!36}$0.36$ & \cellcolor{negcolor!9}$-0.09$ & \cellcolor{poscolor!48}$0.48$ & \cellcolor{poscolor!10}$0.10$ & \cellcolor{negcolor!5}$-0.05$ & $1.00$ &&&& \\
$f_{\rm bg}~[\mathrm{MeV}^{-3}]$ & $50\pm69$ & $22\pm64$ & $18\pm61$ & \cellcolor{negcolor!45}$-0.45$ & \cellcolor{poscolor!16}$0.16$ & \cellcolor{negcolor!49}$-0.49$ & \cellcolor{poscolor!10}$0.10$ & \cellcolor{negcolor!37}$-0.37$ & \cellcolor{negcolor!20}$-0.20$ & $1.00$ &&& \\
$a~[\mathrm{GeV}^{-2}]$ & $0.83\pm0.67$ & $0.77\pm0.32$ & $0.75\pm0.32$ & \cellcolor{poscolor!2}$0.02$ & \cellcolor{negcolor!6}$-0.06$ & \cellcolor{poscolor!0}$0.00$ & \cellcolor{negcolor!26}$-0.26$ & \cellcolor{poscolor!42}$0.42$ & \cellcolor{negcolor!72}$-0.72$ & \cellcolor{negcolor!3}$-0.03$ & $1.00$ && \\
$u_{\pm\rho}$ & $-77.6\pm59.4$ & $-30.4\pm9.3$ & $-18.1\pm4.8$ & \cellcolor{poscolor!61}$0.61$ & \cellcolor{negcolor!26}$-0.26$ & \cellcolor{poscolor!74}$0.74$ & \cellcolor{negcolor!32}$-0.32$ & \cellcolor{poscolor!90}$0.90$ & \cellcolor{negcolor!12}$-0.12$ & \cellcolor{negcolor!38}$-0.38$ & \cellcolor{poscolor!60}$0.60$ & $1.00$ & \\
$v_0~[\mathrm{GeV}^{-2}]$ & $0.013\pm0.213$ & $0.036\pm0.025$ & $0.016\pm0.011$ & \cellcolor{poscolor!60}$0.60$ & \cellcolor{negcolor!19}$-0.19$ & \cellcolor{poscolor!70}$0.70$ & \cellcolor{poscolor!13}$0.13$ & \cellcolor{poscolor!33}$0.33$ & \cellcolor{poscolor!6}$0.06$ & \cellcolor{negcolor!36}$-0.36$ & \cellcolor{poscolor!13}$0.13$ & \cellcolor{poscolor!54}$0.54$ & $1.00$ \\
$\chi^2/\mathrm{dof}$ & $0.68$ & $0.66$ & $0.64$ &&&&&&&&&& \\
\bottomrule
\end{tabular}
\end{table}

\begin{table}[htbp]
\caption{The same as Table~\ref{tab:paras-I0} but for Scheme-I1.}
\label{tab:paras-I1}
\renewcommand{\arraystretch}{1.5}
\begin{tabular}{lcccrrrrrrrrrr}
\toprule
Parameters & $\Lambda=0.6~\mathrm{GeV}$ & $\Lambda=1.0~\mathrm{GeV}$ & $\Lambda=1.35~\mathrm{GeV}$ & \multicolumn{10}{c}{Correlation matrix ($\Lambda=1.0~\mathrm{GeV}$)} \\
\midrule
$P_0^{\rm B}/10^3~[\mathrm{GeV}^{-3/2}]$ & $4.11\pm1.07$ & $3.30\pm0.78$ & $2.20\pm0.60$ & $1.00$ &&&&&&&&& \\
$P_{\pm}^{\rm B}/P_0^{\rm B}$ & $0.62\pm0.30$ & $0.55\pm0.34$ & $0.61\pm0.48$ & \cellcolor{negcolor!78}$-0.78$ & $1.00$ &&&&&&&& \\
$P_0^{\rm L}/10^3~[\mathrm{GeV}^{-3/2}]$ & $7.45\pm1.61$ & $6.00\pm1.06$ & $3.98\pm0.67$ & \cellcolor{poscolor!78}$0.78$ & \cellcolor{negcolor!33}$-0.33$ & $1.00$ &&&&&&& \\
$P_{\pm}^{\rm L}/P_0^{\rm L}$ & $0.53\pm0.08$ & $0.518\pm0.049$ & $0.62\pm0.08$ & \cellcolor{negcolor!29}$-0.29$ & \cellcolor{poscolor!17}$0.17$ & \cellcolor{negcolor!43}$-0.43$ & $1.00$ &&&&&& \\
$C_{0X}~[\mathrm{GeV}^{-2}]$ & $-12.57\pm0.78$ & $-3.58\pm0.11$ & $0.254\pm0.043$ & \cellcolor{poscolor!52}$0.52$ & \cellcolor{negcolor!28}$-0.28$ & \cellcolor{poscolor!66}$0.66$ & \cellcolor{negcolor!59}$-0.59$ & $1.00$ &&&&& \\
$C_{1X}~[\mathrm{GeV}^{-2}]$ & $-17.35\pm1.13$ & $-11.26\pm0.30$ & $-8.28\pm0.15$ & \cellcolor{poscolor!22}$0.22$ & \cellcolor{negcolor!4}$-0.04$ & \cellcolor{poscolor!33}$0.33$ & \cellcolor{poscolor!13}$0.13$ & \cellcolor{negcolor!25}$-0.25$ & $1.00$ &&&& \\
$f_{\rm bg}~[\mathrm{MeV}^{-3}]$ & $52\pm61$ & $17\pm63$ & $18\pm61$ & \cellcolor{negcolor!43}$-0.43$ & \cellcolor{poscolor!18}$0.18$ & \cellcolor{negcolor!46}$-0.46$ & \cellcolor{poscolor!17}$0.17$ & \cellcolor{negcolor!33}$-0.33$ & \cellcolor{negcolor!11}$-0.11$ & $1.00$ &&& \\
$a~[\mathrm{GeV}^{-2}]$ & $0.82\pm0.33$ & $0.80\pm0.36$ & $0.78\pm0.33$ & \cellcolor{poscolor!6}$0.06$ & \cellcolor{negcolor!8}$-0.08$ & \cellcolor{poscolor!5}$0.05$ & \cellcolor{negcolor!30}$-0.30$ & \cellcolor{poscolor!51}$0.51$ & \cellcolor{negcolor!77}$-0.77$ & \cellcolor{negcolor!6}$-0.06$ & $1.00$ && \\
$u_{\pm\rho}$ & $-79.7\pm29.7$ & $-28.8\pm8.9$ & $-17.7\pm4.8$ & \cellcolor{poscolor!55}$0.55$ & \cellcolor{negcolor!28}$-0.28$ & \cellcolor{poscolor!68}$0.68$ & \cellcolor{negcolor!43}$-0.43$ & \cellcolor{poscolor!88}$0.88$ & \cellcolor{negcolor!34}$-0.34$ & \cellcolor{negcolor!35}$-0.35$ & \cellcolor{poscolor!69}$0.69$ & $1.00$ & \\
$v_0~[\mathrm{GeV}^{-2}]$ & $0.025\pm0.835$ & $0.312\pm0.194$ & $0.143\pm0.100$ & \cellcolor{poscolor!53}$0.53$ & \cellcolor{negcolor!19}$-0.19$ & \cellcolor{poscolor!64}$0.64$ & \cellcolor{poscolor!4}$0.04$ & \cellcolor{poscolor!4}$0.04$ & \cellcolor{poscolor!14}$0.14$ & \cellcolor{negcolor!31}$-0.31$ & \cellcolor{negcolor!6}$-0.06$ & \cellcolor{poscolor!30}$0.30$ & $1.00$ \\
$\chi^2/\mathrm{dof}$ & $0.68$ & $0.66$ & $0.64$ &&&&&&&&&& \\
\bottomrule
\end{tabular}
\end{table}

\begin{figure}
    \centering
    \includegraphics[width=\linewidth]{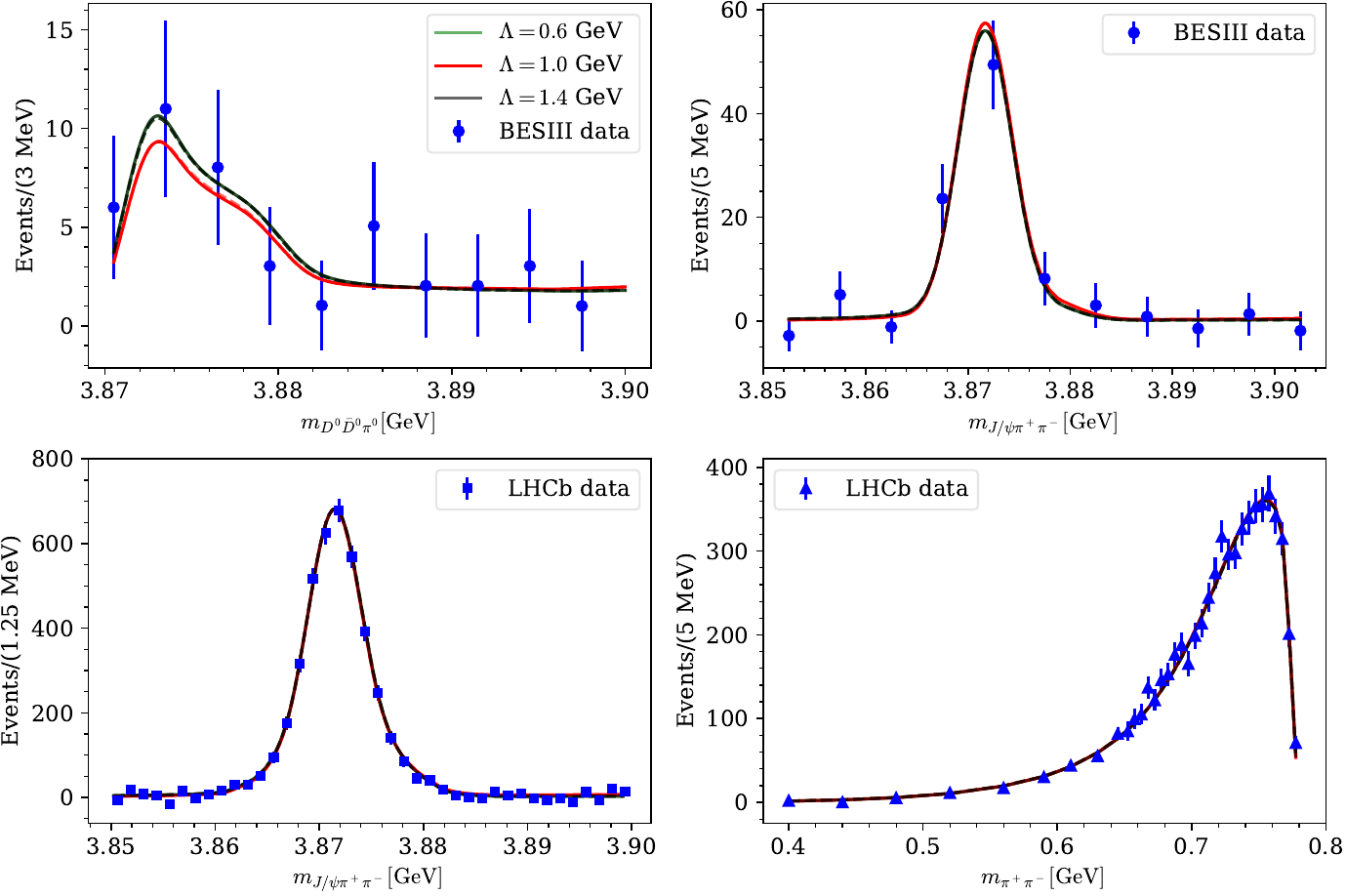}
    \caption{Comparison between the best fitted lineshapes with \(\Lambda=0.6,\ 1.0,\) and
\(1.4~\mathrm{GeV}\) in both Scheme-I0 (solid) and Scheme-I1 (dashed). The
lineshapes in these two schemes are almost indistinguishable.
Panel descriptions: \textbf{(a)} \(D^0\bar D^0\pi^0\) distribution compared
with the BESIII data. \textbf{(b)} BESIII \(J/\psi\pi^+\pi^-\) distribution.
\textbf{(c)} LHCb \(J/\psi\pi^+\pi^-\) distribution. \textbf{(d)} LHCb
\(\pi^+\pi^-\) distribution.
}

    \label{fig:difflamb}
\end{figure}

\section{Extracting the pole terms of $X(3872)$ and $W_{c1}^0$ with Flatt\'e parameterization}\label{app:flatte}
The couplings of the two poles to elastic channels, which can be extracted by $g_{\alpha}g_{\beta}=\lim_{E\to E_{\rm pole}}(E-E_{\rm pole})T_{\alpha\beta},$ read
\begin{equation}\label{eq:gs}{
\begin{aligned}
g_{X,0}&=(0.35\pm0.03)e^{0.12^{+0.08}_{-0.04}\,i}~{\rm GeV}^{-\frac{1}{2}},\\
 g_{X,\pm}&=(0.22\pm0.02)e^{0.09^{+0.09}_{-0.05}\,i}~{\rm GeV}^{-\frac{1}{2}},\\
     g_{W,0}&=(0.44\pm0.07)e^{-1.1^{+0.05}_{-0.09}\,i}~{\rm GeV}^{-\frac{1}{2}},\\
    g_{W,\pm}&=\left(0.70^{+0.03}_{-0.06}\right)e^{-2.18^{+0.16}_{-0.05}\,i}~{\rm GeV}^{-\frac{1}{2}}.
\end{aligned}}
\end{equation}
One notices that both the $X(3872)$ and the $W_{c1}$ have stronger isospin-0 couplings. The two poles we are referring to lie close to the physical axis and thus leave visible imprints on the lineshapes, while their shadow poles are far from the physical region and have little phenomenological impact. Interestingly, the shadow pole of the $X(3872)$ is dominated by the isospin-1 component, indicating significant isospin breaking in these near-threshold states. 
Since charged $W_{c1}^{\pm}$ partners are predicted in the same isospin triplet and also appear as threshold cusps~\cite{Zhang:2024fxy}, we identify the pole on RS${+-}$ discussed in the main text as the neutral member $W_{c1}^{0}$.

To investigate the contributions of the \( X(3872) \) and \( W_{c1}^0 \) to the event distributions, we use the Flatté parameterization with its real parameters adjusted to reproduce both pole locations and residues
\begin{align}
    T_{\alpha\beta}^{\text{Flatté}} = \frac{1}{4\sqrt{m_{\alpha1}{m_{\alpha2}m_{\beta1}m_{\beta2}}}} \frac{\tilde g_{\alpha}\tilde g_{\beta}}{z - \tilde m^2 + \tilde g_0^2 \Sigma_0(z) + \tilde g_\pm^2 \Sigma_\pm(z) +i\, \tilde m\tilde \Gamma}, \label{eq:flatte}
\end{align}
where $\sqrt z=E+m_{D^{0}}+m_{\bar D^{*0}}$ is the total energy in c.m. frame, \begin{align}
    \Sigma_\alpha(z)=\frac{z-z_{\alpha0}}{\pi}\int_{z_{\rm th, \alpha}}^\infty \frac{dz'}{(z'-z_{\alpha0})(z'-z)} \frac{k_{\alpha}(z')}{8\pi \sqrt{z'}} 
\end{align} 
is the self-energy function of channel-$\alpha$ with $k_{\alpha}(z)$ the on-shell momentum, once subtracted at $z_{\alpha0}=z_{\rm th, \alpha}=(m_{\alpha1}+m_{\alpha2})^2$. The bare parameters, \( \tilde g_{\alpha} \), \( \tilde m \), and \( \tilde \Gamma \), are adjusted to reproduce the pole positions and couplings to the \( D^0\bar{D}^{*0} \) and \( D^+D^{*-} \) channels. The prefactor is introduced to correct for the different
 normalization of the non-relativistic $T_{\alpha\beta}^{\text{Flatté}}$. For \( X(3872) \), the parameters are
\begin{align}
    \tilde g_{0} &= 44.3\ {\rm GeV}\,, \quad \tilde g_{\pm} = 27.5\ {\rm GeV}\,, \quad \tilde m = 3.698~{\rm GeV} \,, \quad \tilde \Gamma = 29~{\rm MeV}\, \quad \text{for Scheme-I0},\\
    \tilde g_{0} &= 35.9\ {\rm GeV}\,, \quad \tilde g_{\pm} = 22.1\ {\rm GeV}\,, \quad \tilde m = 3.759~{\rm GeV} \,, \quad \tilde \Gamma = 19~{\rm MeV}\, \quad \text{for Scheme-I1}.
\end{align}

It turns out that for the $W_{c1}$, we cannot reproduce the pole position and couplings using the analogous Flatt\'e parametrization. Therefore, in this case a background term would need to be introduced as outlined in Ref.~\cite{Heuser:2024biq}.  
 Alternatively, we define $T_{\alpha\beta}^{W}=T_{\alpha\beta}-T_{\alpha\beta}^{\text{Flatt\'e},X}$ to calculate the $W_{c1}$ contributions.
Using the parameters above and replacing \( T_{\alpha\beta} \) with $T_{\alpha\beta}^{\text{Flatt\'e},X}$ or $T_{\alpha}^{W}$, we obtain the contributions of \( X(3872) \) and \( W_{c1}^0 \) to the event distributions, shown as the green dashed and gray dash-dotted curves in Figs.~3 and 5 in the main text, respectively.
We also find that using only the $X(3872)$ pole term from a Laurent expansion of the scattering amplitudes, 
which takes the form of a fixed width Breit-Wigner function, instead of the Flatt\'e parameterization in Eq.~\eqref{eq:flatte}, the $X(3872)$ contribution in the $D^0\bar D^0\pi^0$ final state would be significantly underestimated.
\section{ Estimate of the $W_{c1}^0$ significance}\label{app:Wsignificance}
Before presenting results switching off the isospin-1 interactions, let us comment on the fits performed by the LHCb~\cite{LHCb:2020xds,LHCb:2022jez} and BESIII~\cite{BESIII:2023hml} Collaborations. In Refs.~\cite{LHCb:2020xds,BESIII:2023hml}, the $X(3872)$ is parameterized using a Flatt\'e amplitude, where the couplings of $X(3872)$ to $D^0\bar D^{*0}$ and $D^+D^{*-}$ are taken to be equal ($g_{X,0}=g_{X,\pm}$), i.e., assuming the $X(3872)$ to be a pure isoscalar state. 
If we perform a simultaneous fit to both the LHCb and BESIII data following this ansatz, we obtain a compositeness of the $X(3872)$ of about $95\%$, indicating a dominant molecular component.
Moreover,
for the $\pi\pi$ distribution, the LHCb analysis in Ref.~\cite{LHCb:2022jez} introduced a coupled-channel $K$-matrix description of the $2\pi$–$3\pi$ scattering and allowed $X(3872)$ to couple to the $2\pi$ and $3\pi$ channels with coupling strengths characterized by two free parameters, $\alpha_{2\pi}$ and $\alpha_{3\pi}$. These two parameters are essential for determining the isospin breaking in the $X(3872)$ decays, and the isospin-breaking ratio was found to be $R_X=0.29(4)$. 
This ratio is anomalously large compared to the typical isospin-breaking ratio at the amplitude level, which is of $\mathcal{O}((m_d-m_u)/\Lambda_{\rm QCD})\sim\mathcal{O}(\alpha)\sim \mathcal{O}(1\%)$. 

As a dominantly molecular state, the decay dynamics of the $X(3872)$, including the large isospin breaking, are driven by the interactions in the $D\bar D^*$ channels. Due to the non-vanishing values of  the splittings between the charged and neutral $D^{(*)}$ mesons, the isospin-1 and isospin-0 $D\bar D^*$ channels are coupled to each other, and the presence of the isospin-1 $D\bar D^*$ component determines the magnitude of isospin breaking in the $X(3872)$ decays, and vice versa.
It is this that drives the sizable $R_X$ that in
this way leads to the emergence of the isovector $W_{c1}$ in our chiral effective field theory framework, which includes charged and neutral $D$, $\bar D$, $D^*$, $\bar D^*$, and $\pi$ mesons as the degrees of freedom and is the minimal framework to describe the $X(3872)$
as a molecular state.

In the full model we cannot simply remove the $W_{c1}$ from the amplitude. Thus,
to demonstrate how the results change when the $W_{c1}^0$ contribution is switched off, we adopt two simplified alternative models in which the one-pion exchange (OPE) potential and the inelastic potential are turned off. The parameters of the contact potential $V^{\rm ct}$, namely $C_{0X}$ and $C_{1X}$, correspond to the isoscalar and isovector interactions, respectively. We perform two additional fits: one with $C_{1X}$ fixed to zero and the other with $C_{1X}$ left free. The results of the following three schemes are summarized in Table~\ref{tab:comp}:
\begin{itemize}%[nosep,leftmargin=*]
    \item Scheme A: Potential contains only $V_{\rm ct}$ with fixed $C_{1X}=0$;
    \item Scheme B: Potential contains only $V_{\rm ct}$ with free $C_{1X}$;
    \item Scheme C: The one reported in the main text where the potential contains all contributions and $C_{1X}$ is free.
\end{itemize}
\begin{table*}[tb]
    \caption{{Comparison of different fit schemes. Here, $N_p$ is the number of parameters, and the $X(3872)$ and $W_{c1}$ poles are defined relative to the $D^0\bar D^{*0}$ and $D^+D^{*-}$ thresholds, respectively. Only central values are shown here.}}\label{tab:comp}
    \renewcommand{\arraystretch}{1.4}
    \centering
{%
\begin{tabular}{lcccccccc}
\toprule
 Scheme &\ \ $\chi^2$\ \ \ &\ \  $N_p$\ \ \ &\ $\ \chi^2/$dof\ \ \ &\ $X(3872)$ pole (keV)\ \ & \ $W_{c1}$ pole (MeV)\ \ &\ $R_X$\ \  &\ AIC\ \ &\ BIC\ \ \\ \midrule
 A (no OPE, $C_{1X}=0$) &$474$ & 8 & $5.4$ & $-298-15i$ &  $-$ & $0.013$ & $490$ & $511$ \\
  B (no OPE, $C_{1X}$ free) &$83$ & 9 & $0.95$ & $-125-20i$&  $1.0+0.8i$ & $0.26$ & $101$ & $124$ \\
  C (full framework as in main text)  &$57$ & 10 & $0.66$ & $-160-125i$ &  $3.7+1.3i$ & $0.26$ & $77$ & $102$ \\ \bottomrule
\end{tabular}}
\end{table*}

\begin{figure}[!t]
    \vspace{10pt}
    \includegraphics[width=0.8\linewidth]{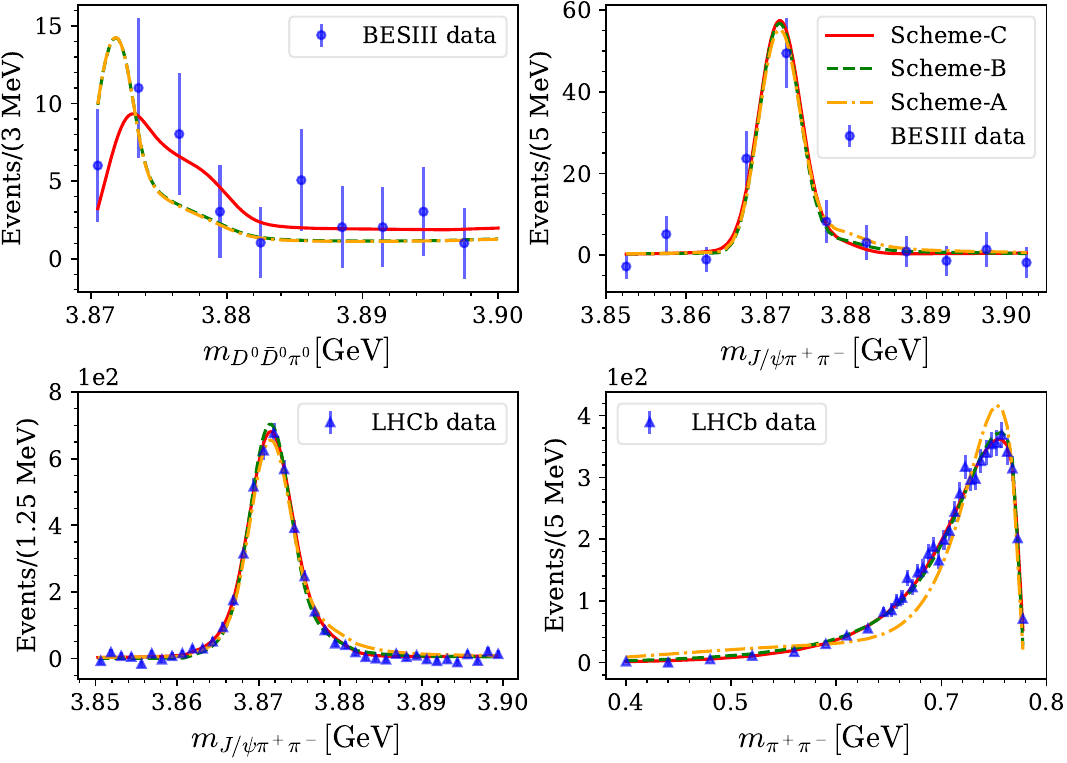}
    \caption{Comparison of the best fitted lineshapes in different schemes. Panel descriptions: \textbf{(a)} \(D^0\bar D^0\pi^0\) distribution compared
with the BESIII data. \textbf{(b)} BESIII \(J/\psi\pi^+\pi^-\) distribution.
\textbf{(c)} LHCb \(J/\psi\pi^+\pi^-\) distribution. \textbf{(d)} LHCb
\(\pi^+\pi^-\) distribution.}
    \label{fig:simplifiedmodel}
\end{figure}

A comparison of the best-fitted lineshapes for the three schemes is shown in Fig.~\ref{fig:simplifiedmodel}. Table~\ref{tab:comp} highlights three main points. First, the nonvanishing $R_X$ in scheme A is of $\mathcal O(1\%)$, as expected. 
Second, both our full framework (scheme C) and the simplified model with $C_{1X}$ (scheme B) provide a good description of the data---both contain a $W_{c1}$ pole. Third, the parameter $C_{1X}$ plays a crucial role in improving the fit quality, particularly for the $\pi\pi$ distribution, as expected. In the simplified model, the significance of $W_{c1}^0$ reaches $\sqrt{474-83}~\sigma\approx 20~\sigma$ (based on statistical uncertainties only), far exceeding the standard $5\sigma$ threshold. This exceptionally high significance is clearly reflected in the $\pi\pi$ invariant mass distribution, as evidenced by the extremely poor fit quality of the orange curve in Fig.~\ref{fig:simplifiedmodel} from scheme A. Scheme B is sufficient to account for the isospin breaking of $X(3872)$ and the existence of $W_{c1}^0$. However, scheme B does not have the full $D\bar D\pi$ channel contribution (the $D^*$ self-energy is kept while the OPE is switched off), 
violating three-body unitarity, and therefore cannot properly capture the $D\bar D\pi$ dynamics that dominates the $X(3872)$ width. 

To further assess the trade-off between goodness-of-fit and model complexity from a statistical perspective, we calculate the Akaike Information Criterion (AIC)~\cite{Akaike:1974vps} and the Bayesian Information Criterion (BIC)~\cite{Schwarz1978BIC}. These criteria evaluate models for having more parameters:
\begin{itemize}%[nosep,leftmargin=*]
    \item AIC is defined as $2 N_p+\chi^2$, where $N_p$ is the number of parameters. It balances the fit quality ($\chi^2$) against model complexity.
    \item BIC is defined as $N_p \log N+\chi^2$, where $N$ is the number of data points. BIC imposes a stronger penalty for additional parameters, especially when the dataset is large.
\end{itemize}
For scheme C, both AIC and BIC are significantly lower than those for scheme B, indicating that the additional OPE and inelastic potentials improve the fit quality while remaining statistically justified.
\end{appendix}
\end{onecolumngrid}
\end{document}